\documentclass[11pt, twoside, letter]{article}
\usepackage[margin = 1in]{geometry}
\usepackage[utf8]{inputenc}
\usepackage{amsmath, amssymb, amsfonts, cite, authblk}
\usepackage{ifthen}
\usepackage[boxed, linesnumbered]{algorithm2e}
\usepackage{comment}
\usepackage{url}

\usepackage{pgf}
\usepackage{tikz}
\usetikzlibrary{arrows,automata}

\usepackage{microtype}

\usepackage{amsthm}
\usepackage{hyperref}
\usepackage{multirow, multicol}
\usepackage{subfigure, graphicx} 
\usetikzlibrary{arrows,shapes}

\tikzset{circle/.pic={
\node[circle, aspect=1, draw, minimum size=0.3cm, text width=0.2cm] () at (0,0) {\tikzpictext};
}} 
\usepackage{natbib}
\usepackage{microtype}
\usepackage{pgfplots}
\usepackage{booktabs}
\pgfplotsset{compat=1.10}

\bibpunct{[}{]}{;}{a}{,}{;}
\begin{document}

\title{How bad is selfish routing in practice?}
\date{}

\author[]{Barnab\'e Monnot \; Francisco Benita \; Georgios Piliouras \\
Singapore University of Technology and Design \\
\texttt{monnot\_barnabe@mymail.sutd.edu.sg, \{francisco\_benita,georgios\}@sutd.edu.sg}}

\maketitle

\begin{abstract}
Routing games are one of the most successful domains of application of game theory. It is well understood
that simple dynamics converge to equilibria, whose performance is nearly optimal regardless of the
size of the  network or the number of agents.
These strong theoretical assertions prompt a natural question: How well do these
pen-and-paper calculations agree with the reality of everyday traffic routing?  
We focus on a semantically rich  dataset from Singapore's National Science Experiment that captures detailed information about the daily behavior of thousands of Singaporean
students. Using this dataset, we can identify the routes as well as  the modes of transportation used by the students, e.g. car (driving or being driven to school) versus bus or metro, estimate
source and sink destinations (home-school) and trip duration,  as well as their mode-dependent available routes.
We quantify both the system and individual optimality.
Our estimate of the Empirical Price of Anarchy lies between $1.11$ and $1.22$.  Individually, the typical behavior is consistent from day to day and nearly optimal, with low regret for not deviating to alternative paths.

\end{abstract}

\section{Introduction}
Congestion games are amongst the most historic, influential and well-studied classes of games.
Naturally, as the name suggests, they were designed to capture settings where the payoff of each agent depends on the resources he chooses and how congested each of them is.
Proposed in \citep{rosenthal73} and isomorphic to potential games \citep{potgames}, they have been successfully employed in a myriad of modeling problems. Naturally, one application stands above the rest: modeling traffic. Having strategy sets correspond to the possible paths between source and sink nodes in a network is such a mild and intuitive restriction that routing/congestion games are effectively synonymous to each other and jointly mark a key contribution of the field.

Routing games have also played a seminal role in the emergence of algorithmic game theory.
The central notion of Price of Anarchy (PoA), capturing the inefficiency of worst case equilibria, was famously first introduced and analyzed in routing games \citep{KoutsoupiasP99WorstCE,roughgarden2002bad}. Routing games have set the stage for major developments in the area such as the introduction of regret-minimizing agents 
\citep{pota}  that eventually led to the consolidation of most known results under the
umbrella of  $(\lambda,\mu)$-smoothness arguments \citep{Roughgarden09}. Impressively, this work
also established that these guarantees are robust for a wide variety of solution concepts such as correlated equilibria,
regret-minimizing agents and approximations thereof.
Finally, congestion games still drive innovation in the area with results which extend the strength and  applicability of PoA bounds for large routing games  \citep{feldman2016price}, or even to dynamic populations
 \citep{lykouris2016learning}.

With every successive analytical achievement seemingly chopping slowly away the distance between theoretical models and everyday reality, the PoA constants for routing games, e.g., the $4/3$ for the nonatomic linear case \citep{roughgarden2002bad} have become something akin to the universal constants of the field.
Small, concise, dimensionless, they seem almost by their very nature to project purity and truth.
But do they? After all, there are many of them. In the case of quadratic cost functions PoA
$\approx1.626$,
 whereas for quartic functions, which have been proposed as a reasonable model of road traffic, PoA
 $\approx 2.151$ \citep{Sheffi1985urban,RoughgardenBook16}. Do these ``back-of-the-envelope'' theoretical calculations have any predictive power \textit{in practice}?

\smallskip

 \textit{Our goal} is to perform the first-to-our-knowledge experimental investigation of PoA in large scale real world traffic networks. Secondly, we wish to estimate inefficiencies at the individual level by quantifying the daily empirical regret for each agent, i.e., how much faster could each agent have reached their destination if they had clairvoyant access to all the traffic information and made the best possible routing decision. Finally,
 we wish to quantify the extent to which the system has reached stasis, i.e., what percentage of users choose the same route from one day to the next.

  \smallskip

Let's portray the task at hand via an example. Suppose Alice is a commuter in our traffic network.
Since we wish to understand the optimality of her route from her perspective, i.e. her regret (or lack thereof) for not having chosen her optimal path,  we need to track Alice's motion
 and not merely gather anonymous traffic flow information.
 Naturally, we wish to compute society-wide statistics such as the system efficiency and the average/median regret as well as statistics on the consistency of routing decisions. 
Thus, we need to gather movement data at the individual level for a large number of commuters.  Gathering and processing routing information for all commuters is practically infeasible, therefore, we need to isolate a sample of commuters large enough and diverse enough in its routing needs so as to derive a meaningful approximation of the true statistics.

 Our next challenge is to abstract away from the noisy reality of movement information the building blocks necessary for applying game-theoretic rationality:  Which are Alice's source and sink destination?
 Which are Alice's available paths? This depends on whether Alice is using a car or public transportation.
 In fact, this information is critical to computing regret. Indeed, assuming Alice uses public transit, then she should not be penalized for not using a shorter route only accessible by car. Since we are gathering information on a large number of users we cannot incentivize them to consistently report this information. Hence, we need to derive it implicitly from the data.

Computing the empirical regret or the PoA involves answering hypothetical what-if questions.
 ``What if I had chosen route $B$ today?''  Alice wonders stuck in traffic along highway $A$. If in the sample population there exists another user, Bob, a neighbor of Alice, who has to commute to the same destination at the same time with her then we can access this hypothetical state. To employ this kind of techniques we need to identify the clusters of comparable users and test for robustness to different clustering methods. Finally, we need to bound the optimal social cost and run these experiments repeatedly on consecutive days to check whether commuters are typically exploring alternate routes or have settled on a preferred route.

 The list of challenges above, which is by no means complete, suggests  that our  goal is tricky.
 On the other hand, one can readily point out to dozens, if not hundreds, of theoretical papers successfully analyzing the very questions that we pose here bypassing all of these obstacles. 
  Is an experimental investigation really worth the effort? We believe that it is and the reasons are threefold.

First, the abundance of theoretical inquiries naturally invites the need for experimental validation.
The more successful a theoretical model, the more pressing the requirement for testing.
The belief that routing networks in practice have a small PoA is arguably entrenched within our community for more than a decade and thus we see this paper as delivering a much belated check.
Second, the PoA results themselves, even if they turn out to be validated by the data, are by their very nature overly pessimistic. Experimental estimates can provide us
with more accurate estimates about  the inefficiencies of typical instances and possibly help us refine our theoretical models as well.
Finally, the performance metrics  developed and refined within algorithmic game theory (such as PoA or regret) are remarkably concise and conceptually simple whilst at the same time encapsulating key performance indicators. Estimating these metrics, which are not typical from a transportation literature perspective~\citep{Sheffi1985urban}, could help guide public policy decisions.  

\subsection*{Summary of results}

\begin{itemize}
    \item We estimate that the Empirical Price of Anarchy in Singapore lies between 1.11 and 1.22 overall, with marked contrast when discriminating by mode of transportation (see section \ref{sec:poa}).
    \item We compute the empirical regret distribution and show it has a median value of 4 minutes 40 seconds and mean approaching 6 minutes (see section \ref{sec:regret}).
    \item Finally, we show that most students use the same means of transportation across trips and that a large number of them consistently selects the same route (see section \ref{sec:consistency}).
\end{itemize}

\subsection*{Methodology in a nutshell}

We focus on a semantically rich dataset from Singapore's National Science Experiment (NSE), a nationwide ongoing educational initiative led by researchers from the Singapore University of Technology and Design (SUTD)~\citep{Monnot2016,Wilhelm2016}.  This dataset includes detailed information about the daily behavior of tens of thousands of Singapore students that carry custom-made sensors
resulting in millions of measurements. The students are dispersed throughout the city-state and their daily commutes to school are reasonably long for them to meaningfully interact and experience the daily traffic.

From this dataset, we can identify the mode of transportation used by the students, e.g. car (driving or being driven to school) versus bus or metro, estimate source and sink destinations (focusing on home-school pairs) as well as their mode-dependent available routes. By exploiting this effectively synchronous movement\footnote{Neighboring students that use the same mode of transportation and go to the same school will typically head out at around the same time.}, we can quantify the optimality of individual decisions (e.g. are agents using the best path available to them?)  as well as the efficiency of the network at the social cost level. See section \ref{sec:methodology} for a detailed breakdown of the implementation and methodological approach including a description of the experiment, the datasets, as well as the algorithmic techniques deployed.

\section{Findings}
To analyze the equilibrium properties of the routing network used by Singapore students, we focus on three different measures:
\begin{itemize}
    \item \textbf{Optimality:} How does the trip of one student compare with the optimal directions?
    \item \textbf{Regret:} How does the trip duration of one student compare with those around her?
    \item \textbf{Consistency:} Do students use the same routes to reach the school?
\end{itemize}

\subsection{Societal Optimality and Empirical Price of Anarchy}
\label{sec:poa}

In this section, an approximation of the Price of Anarchy for the Singapore students using the transportation network is derived. Price of Anarchy was introduced by \citep{KoutsoupiasP99WorstCE} to measure the inefficiency of equilibria in routing games, compared with the optimal social state where overall congestion is minimized. Several tight bounds are known \citep{RoughgardenBook16}, indicating how far the equilibrium is from optimal when different edge latency functions are considered.

Our findings show that it is possible to estimate the \textbf{Empirical Price of Anarchy} by bounding it from above and below. The numerical results point to a value  that is in good agreement with  theoretical predictions. We first define the Empirical PoA (or EPoA) as:
\[
    \texttt{\textbf{EPoA}} = \frac{\texttt{\textbf{Cost}(Recorded trip duration)}}{\texttt{\textbf{Cost}(Optimal trip durations)}}
\]

We use an online trip optimization tool, Google Directions, to collect the optimal trip durations for the students' morning journey to school. These directions are queried in different time periods to capture both light and heavy traffic conditions --- more details are given in our Methodology section.

The numerator is obtained by adding up the duration of all the morning trips of our dataset. We separate users into two categories: the car users and the public transport users. For car users, we sum over the trips for which optimal duration time was recorded in both heavy traffic conditions and light traffic conditions. Additionally, we sum over transit users for which the optimal transit directions were found.

We derive the following upper and lower bounds:
\[
\begin{split}
    \frac{\texttt{\textbf{Cost}(Recorded trip duration)}}{\texttt{\textbf{Cost}(Optimal trip durations (heavy traffic))}} \leq \texttt{\textbf{EPoA}} \\
    \leq \frac{\texttt{\textbf{Cost}(Recorded trip duration)}}{\texttt{\textbf{Cost}(Optimal trip durations (light traffic))}}
\end{split}
\]
Our results show that \( 1.11 \leq \texttt{\textbf{EPoA}} \leq 1.22 \), when the EPoA is computed with both car and transit users.

Discriminating between the two yields a much more contrasted picture. Indeed, the EPoA for transit users is found to be roughly equal to 1, indicating that students using public transportation have little room to improve their trip duration. Conversely, the EPoA varies significantly depending on the traffic conditions for subjects taking private transportation to school. In light traffic conditions, we find EPoA = 1.94, while in heavy traffic (obtained by taking the maximum reported time by Google Directions API), the EPoA falls to 1.40.

Taken as whole, these measures show remarkable proximity to the various PoA results found in the literature, such as the \( 4/3 \) ratio of \citep{roughgarden2002bad} in the case of linear cost functions or the results in \citep{Roughgarden_class} concerning more general cost functions.

\subsection{Individual Optimality and Empirical Regret}
\label{sec:regret}

To answer the question of individual optimality, we compare the durations of the morning trip for the subjects. A fair comparison is only achieved when looking at students leaving from the same neighborhood on the same day and at roughly the same time, going to the same school and using the same mode of transportation. The notion of neighborhood is expanded upon in our following methodology section, where we describe how the clustering of the data was achieved.

In the cases where the class of comparable subjects has more than two individual students, we collect the \textbf{empirical regret} encountered by every student in the class. To do so, we find the student in the class with minimal trip duration and set her regret to zero. For other members of the class, the empirical regret is equal to the (non-negative) difference between their trip duration and the minimal trip duration.

Our notion of empirical regret shares its name with the traditional regret measure, commonly found in the learning and multi-agent systems literature, for the following reason. The players here are faced with multiple strategies that they can choose from: the routes that go from their neighborhood to the destination. They may not know about current traffic conditions or which route will take the least amount of time but nevertheless have to make a decision. A posteriori, this decision can be compared with the best action they could have implemented on that day, and the difference is the regret.

The measure of empirical regret depends naturally on the geographical area covered by the neighborhood. As the area increases, so does the accumulated regret, since the minimum is taken over a larger set of students. However, neighborhoods that are too large lose in precision, as two different subjects in the same cluster may have very different trip lengths. The results in this section use a geographical cluster size of about 400 meters, while we perform sensitivity analysis in the methodology to show the robustness of our findings.

Low empirical regret is a necessary condition for equilibrium. Indeed, at equilibrium, all comparable subjects should perform their trip in roughly the same amount of time. If one individual encounters a regret of say, 10 minutes, she may be better off by switching to a different route, e.g. the one used by the fastest individual in the cluster.

On the other hand, a high empirical regret warns us that some users are unable to find the fastest route to reach their destination. We see two possible directions to explore after such a conclusion. If we assume that individuals are solely interested in minimizing their trip duration --- perhaps a fair assumption for the morning trip, constrained by the hard deadline of the class start ---, then the network may benefit from the injection of information on how to traverse it. Otherwise, a high empirical regret reveals that other factors enter into consideration when the student is selecting the route, such as finding the least expensive one, the more climatised one or one that is shared with other students. The additional data collected by the sensor (e.g. temperature, proximity to other sensors) can indeed be articulated to uncover the nature of these factors.

\begin{figure}
    \centering
    \includegraphics[width=0.5\textwidth]{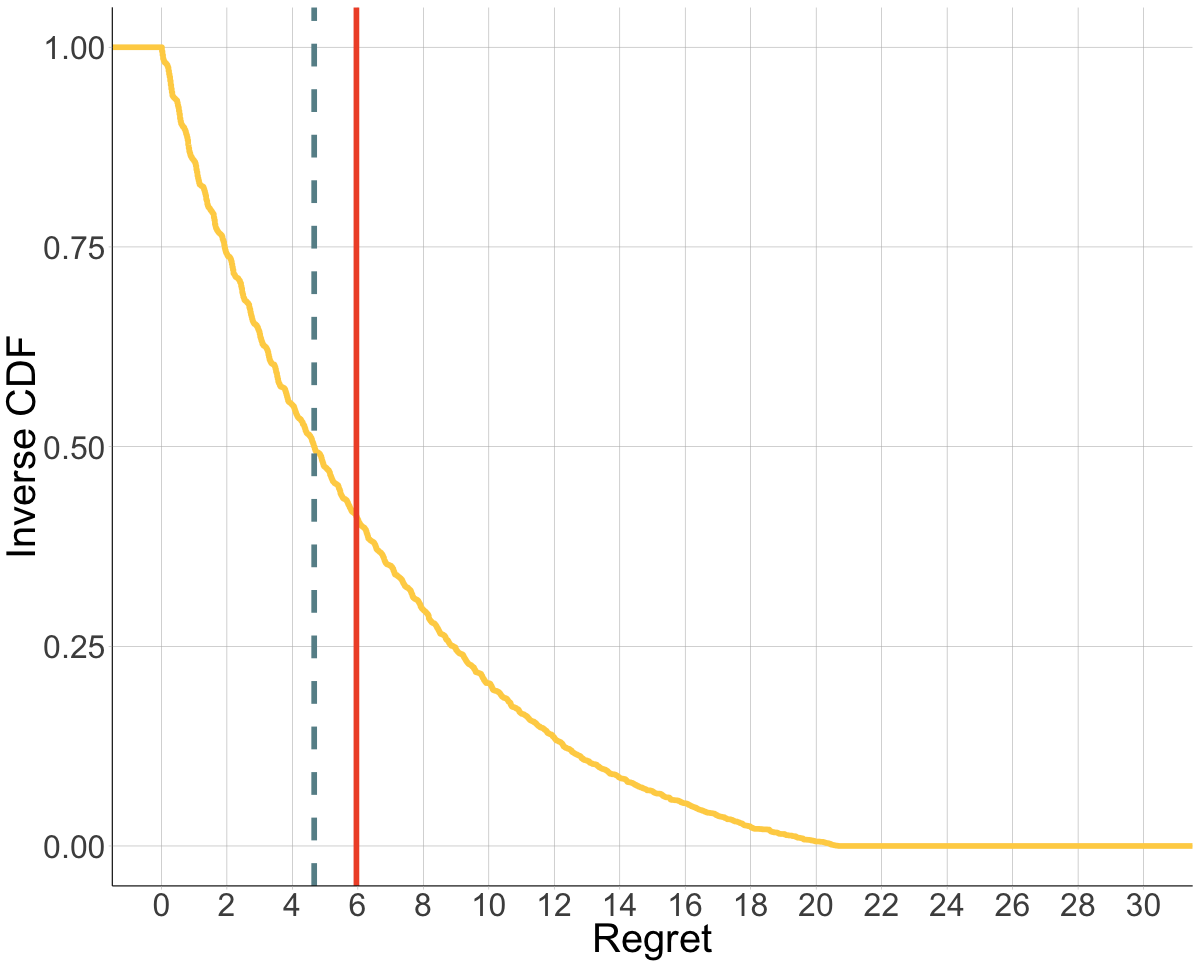}
    \caption{Inverse cumulative distribution function of the regret. We aggregate all days of the experiment in a single figure and remove users who have regret above the 95th percentile or zero regret (in other words, the baseline students). The mean regret signalled by the red line is equal to 6 minutes, while the median regret plotted with the dashed blue line is equal to 4 minutes and 40 seconds.}
    \label{fig:regretcdf}
\end{figure}

In Figure \ref{fig:regretcdf}, we plot the inverse cumulative distribution of the emprirical regret. A point on the curve indicates which fraction of individuals (read on the \(y\)-axis) have empirical regret greater or equal than \(x\) (read on the \(x\)-axis). We also give the mean (solid red line) and median (dashed blue line) experienced empirical regret. It should be noted that the empirical regret distribution and its moments do not include the students for which the regret is zero, i.e. the best in the cluster. Trips with empirical regret above the 95th percentile are also removed, as arbitrarily large regret is sometimes observed due to sensor imprecision.

Larger geographical cluster sizes give rise to larger average empirical regrets, but the results are relatively robust. The mean empirical regret oscillates around 6 minutes, while the median one is situated around 4 minutes and 40 seconds. This result motivates the introduction of a solution parametrised by two values, \( \epsilon \) and \( \delta \). The reported measurements constitute an \( (\epsilon, \delta) \)-equilibrium if we find that a fraction \( 1-\delta \) of users experience at most a quantity \( \epsilon \) of regret. The experiment yields values \( \epsilon = 22 \) minutes and \( \delta = 0.05 \).

\subsection{Equilibration and Empirical Consistency}
\label{sec:consistency}

If we believe that our system is at equilibrium, then we should expect that the students' route decisions do not vary wildly between successive days of study. We investigate the issue from two different angles. First, we compare the modes of transportation selected by one student over the days of the experiment. Second, we improve the previous result by considering whether the selected routes are identical (e.g. always use the same combination of bus and train, or always use the same road on car).

The first analysis shows that more than 60\% of students have used the same mode of transportation in all morning trips available in our dataset. This number arises when we are discriminating between trips where the principal mode of transportation is either the train, the bus or the car. The principal mode is obtained by defining as principal the mode with which the student has made the most distance. The fraction increases to close to two thirds of the samples if we simply separate between the students using public transit from those who use private transportation.

For the second analysis, we have implemented a novel algorithm to determine whether two route choices are identical. We find that for students using the same mode of transportation across all days, the percentage of subjects selecting the same route is very high, in the order of 94\%. We detail in the methodology the algorithm used to obtain this number.

\section{Methodology and Implementation}
\label{sec:methodology}
The following section delves into the details of how the dataset was collected and the implementation of algorithms to work with it. We first give an overview of the National Science Experiment and explain how a large clean dataset was extracted out of the measurements. Second, we expand upon how the raw data was processed into higher-level abstractions such as \textit{trips} and how to compute the mode of transportation. Finally, we give the methodology behind the previous findings supplemented with more detailed results.

\subsection{The Singapore National Science Experiment}

\subsection*{Presentation of the experiment and previous work}

As part of the \citep{SmartNation} initiative, the \emph{National Science Experiment} (NSE) has the primary goal of inspiring future generations of students to pursue technical education. The NSE is a nationwide project in which over 50,000 students from primary, secondary and junior college wore a sensor, called SENSg, for one week in 2015 and 2016. The SENSg sensors collect ambient temperature, relative humidity, atmospheric pressure, light intensity, sound pressure level, and 9-degree of freedom motion data. In \citep{Wilhelm2016}, a detailed technical description of the sensor is presented. The design of the SENSg was created with the specific requirements to have a low-cost device for a one-week crowd-sensing experiment, that does not need to be charged. It led up to the mass-production of 50,000 sensor nodes.

The SENSg scans the Wi-Fi hotspots which are used to localize the sensor nodes as well as to move sensor data to a back-end server. All environment values are sampled every 13 seconds using the Wi-Fi based localization system. The raw collected datapoints are then post-processed to obtain semantic data, employing state-of-the-art methods described in \citep{Monnot2016, Wilhelm2017, Zhou2017}. The semantic data covers the identification of individual trips within the discrete stream of locations --- recapituled in section \ref{sec:tripid} ---, inference of the activity performed at each endpoint and transportation mode classification. We give a brief summary of the latter in section \ref{sec:modeid}.

\subsection*{Overview of Singapore's Routing Network}

\begin{figure}
    \centering
    \includegraphics[width=.8\textwidth]{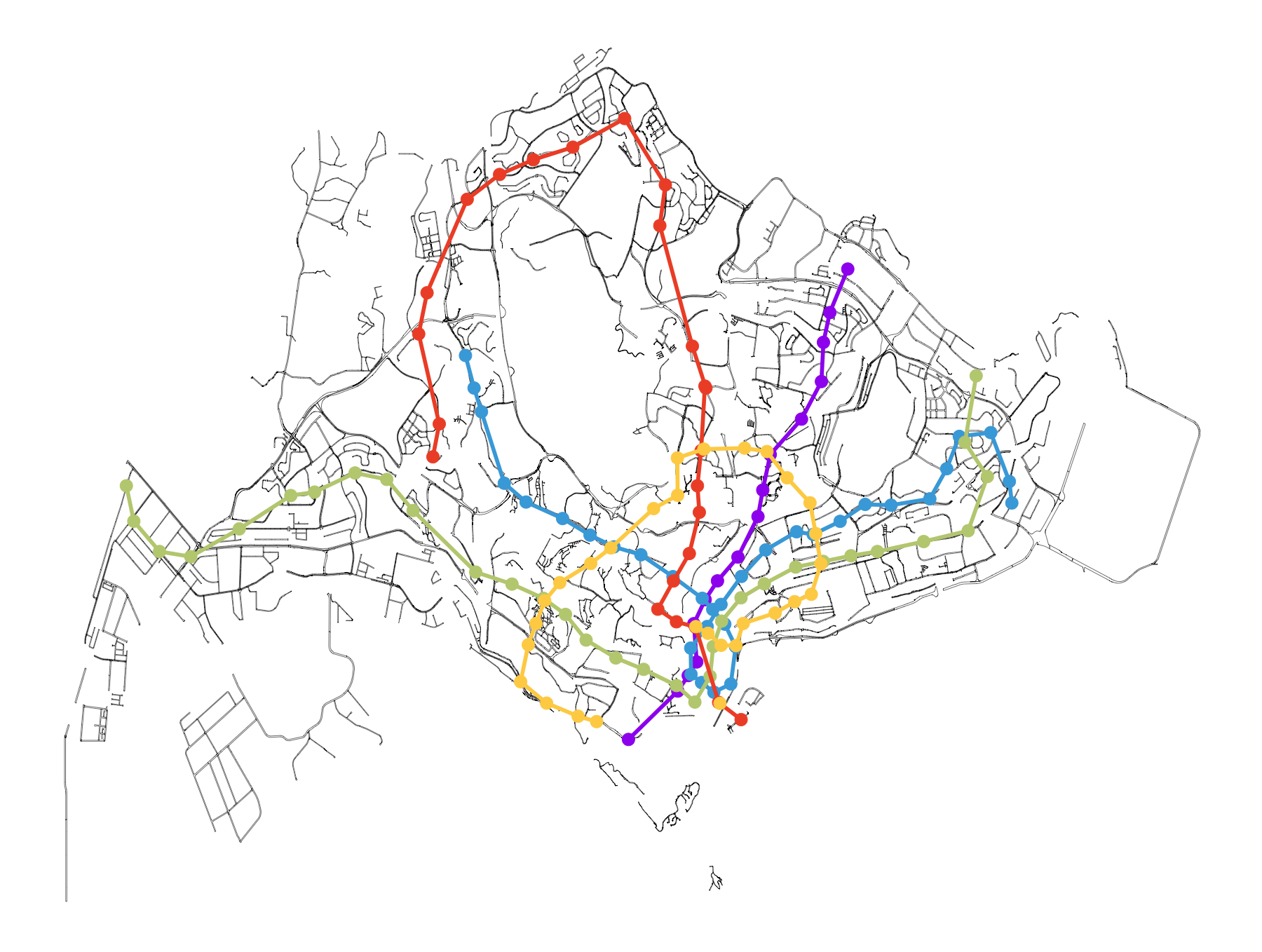}
    \caption{A plan of Singapore with major roads and MRT (Mass Rapid Transit) stations, by line.}
    \label{fig:singaporemap}
\end{figure}

Singapore is an island city-state located at the southern tip of Peninsular Malaysia in South East Asia (depicted in Figure \ref{fig:singaporemap}). According to the General Household Survey 2015 elaborated by the Department of Statistics Singapore, the city-state is home to 5.6 million people, with only an area of 712 km$^{2}$. Metro and bus are the two main modes of public transportation in this densely populated country. Private modes of transportation are mainly passenger cars like Taxis, Ubers and private vehicles.

The 2015 Annual Report published by the Land Transport Authority (LTA) indicates that the metro system consists of a network of 142 stations that can be grouped into 5 Mass Rapid Transit (MRT) lines and 3 Light Rail Transit (LRT) lines. This rail system moves more than 2 million passengers daily, and is the backbone of Singapore's public transport network and spans some 150km across the island. Metro operation hours are from 5:50 am to about midnight daily with a 2 to 3 minutes frequency during peak hours (7 am to 9 am). The studies in \citep{Sun2012, Poonawala2016} make an effort to understand crowds within the Singapore metro network. Singapore's public buses serve local transport within a town to the hub whereas the metro network is used for longer distance trips. The bus network is larger, with 260 bus services and 4,684 bus stops according to the 2015 Annual Report of the LTA. During peak periods, all buses are scheduled to arrive every 15 minutes or less with at least half of these scheduled to arrive every 10 minutes or less.

The metro and bus fares are paid using a contactless smart card, named EZ-Link, where users just tap the card to scan their train ticket at the gantries when commuters enter and exit the metro platform area or bus stop. The LTA regulates and oversees metro and bus transport, ensuring they meet safety standards. The rich information generated by the use of the EZ-Link card was exploited in the works of \citep{Sun2012, Lee2014,Holleczek2015,Poonawala2016}, attempting to find interesting insights for public transit planners.

On the other hand, the rapid development and the population growth of Singapore have led to an increase in the car usage. Still, in a small city-state like Singapore, the expansion of the road network is a constraint. The trade-off between the use of land for roads  (currently accounting for 12\% of the land) and other activities has become an important issue for planning authorities. With this in mind, the Government has put in place a combination of ownership and usage measures to manage road traffic. Private car users have to purchase a paper license that in many cases is more expensive than the car itself. Despite the economic strategy to regulate the car ownership, the 2012 Household Interview Travel Survey found that about 39\% of the trips in 2012 were made by private transport. Moreover, 46\% of households owned cars in 2012, compared to 40\% in 2008, and 38\% in 2004.

\subsection*{Details of the dataset}

The NSE 2016 dataset contains data of 49,526 students who wore the SENSg sensor. This work uses the trip identification algorithm developed in \citep{Zhou2017} where five different modes can be identified, namely: (a) stationary; (b) walking; (c) riding a metro; (d) riding a bus; and (e) riding a car. With additional information from the LTA, the algorithm detection covers 8 metro lines, 106 metro stations, 260 bus services and 4,684 bus stops. Similarly, the 164 km of expressways and the 698 km of arterial roads in Singapore feed the algorithm to distinguish whether a subject is traveling in a car.

The experiment was designed to analyze homogeneous users, i.e. primary, secondary or junior college students, reucing the complexity of understanding mobility patterns. This work focuses primarily on morning travels of students who get to their schools from their homes. Even though the walking mode can be detected by the algorithm, we restrict the study to metro, bus, and car trips. To ensure the quality of our empirical results, we perform a strict data cleaning process, and the top and bottom 5 percents of the trip duration/distance traveled values were dropped to remove outlier points.

Table \ref{tab:dataset} charts the basic description of the dataset used in this study. A total of 32,588 clean trips are considered, covering 15,875 students and 89 schools. The number of students by school type is approximately equally distributed, hence capturing the routing behavior of students over a large space in Singapore. A few pointed insights can be derived from the table. For example, around 55 percent of the Secondary school students take public transport (metro or bus) to reach their schools, and this proportion increases to 75 percent for junior college students. In the case of primary school students, a large fraction of the trips are made by car, a number similar to that of the few adult Singapore residents who use car to reach their workplaces (21.9 percent of the residents according to the LTA reports).

\begin{table}
\centering
\begin{tabular}{lrrrrrr}\toprule
                                & \multicolumn{3}{l}{Mode (\% of the trips)}                       & \multicolumn{3}{c}{Total}                                                              \\ \toprule
\multicolumn{1}{l}{School type} & \multicolumn{1}{c}{Metro} & \multicolumn{1}{c}{Bus} & \multicolumn{1}{c}{Car}   & \multicolumn{1}{c}{Trips} & \multicolumn{1}{c}{Students} & \multicolumn{1}{c}{Schools} \\ \toprule
Primary                         & 5.50                     & 35.20 & 59.30                     & 9,373                     & 4,553 & 28                           \\
Secondary                       & 14.13                     & 40.80                   & 45.07                     & 10,789                    & 5,229 & 20                          \\
Junior College                  & 53.60                     & 22.14                   & 24.26                     & 12,426                    & 6,093                        & 41                          \\ \toprule
                                & \multicolumn{1}{l}{}      & \multicolumn{1}{l}{}    & \multicolumn{1}{r}{\textbf{Total}} & 32,588 & 15,875                       & 89             \\ \toprule
\end{tabular}
\caption{Dataset description}
\label{tab:dataset}

\end{table}

\begin{table}
\centering
\begin{tabular}{lrrrrrr}\toprule
                                & \multicolumn{3}{c}{Average}                               & \multicolumn{3}{c}{Average}                           \\
                                 & \multicolumn{3}{c}{trip duration (mins)}                               & \multicolumn{3}{c}{trip length (km)}                           \\\toprule
\multicolumn{1}{l}{School type} & \multicolumn{1}{c}{Metro} & \multicolumn{1}{c}{Bus} & \multicolumn{1}{c}{Car} & \multicolumn{1}{c}{Metro} & \multicolumn{1}{c}{Bus} & \multicolumn{1}{c}{Car} \\\toprule
Primary                         & 34'06                      & 25'18                    & 22'12                    & 8.9                       & 5.2                     & 7.2                     \\
Secondary                       & 36'24                      & 25'42                    & 22'42                    & 10.5                      & 4.9                     & 7.9                     \\
Junior College                  & 48'54                      & 40'48                    & 39'30                    & 16.8                      & 9.3                     & 14.2           \\ \toprule
\end{tabular}
\caption{Average of the duration and distance of trips}
\label{tab:duration}
\end{table}

Table \ref{tab:duration} presents the average trip duration and the average trip distance by school type and transportation mode. In general, primary school trips cover short distances, as the matching of students to their schools is done on a regional basis by the Education Ministry, so that no child needs to travel long distances to his or her school. On the other hand, the average distance of the secondary and junior college students is in general large, especially for train trips.

\subsection*{Comparable datasets}

We now give an overview of studies comparable to ours:
\begin{itemize}
    \item The works of \citep{Sun2012, Lee2014, Holleczek2015} or \citep{Poonawala2016}, produced in Singapore, are based on smart card data or GSM (cellular phone) data and focus only on metro commuter trips, while we recognize different modes of transportation.
    \item \citep{Holleczek2014} explores the public (bus/metro) and private (taxi) transportation usage as well as the transport mode preferences across Singapore.
    \item Finally, larger studies of Singapore have made use of rich datasets comprising 2.6, 3.4 and 3.9 million users (in \citep{Poonawala2016}, \citep{Holleczek2014} and \citep{Holleczek2015}, respectively), and covering up to 60 million of trips (c.f. \citep{Lee2014}).
\end{itemize}

\subsection{From raw data to semantic data}

\subsection*{Trip Segmentation}
\label{sec:tripid}

We use a rule-based method for the segmentation of geographical location streams into individual trips. A trip is defined as a movement between two Points of Interest (POI), where a POI is marked at a location where the student is spending some amount of time. The trip starts when the student is leaving a POI and ends when she arrives at the destination POI. No student input is necessary for the algorithm to perform the segmentation (e.g. manual labeling of trip endpoints), making it easily scalable to large datasets.

To obtain these POIs, we first denoise the sequence of coordinates by smoothing it with a moving window. A sequence of velocities is computed by estimating the speed between two consecutive locations. When the velocity falls under a fixed threshold, we consider that the student is stopped at the location. If the stop lasts for a duration larger than a second threshold, we mark the location as a POI and end the trip. When the velocity picks up again and the student moves away from the POI, we start a new trip.

Further details of the trip segmentation algorithm are given in \citep{Monnot2016}. For the current study, we collect a dataset of trips joining the student's home to her school in the morning. The home and school locations are also obtained without active reporting by the students and is done by aggregating locations recorded within certain timeframes (e.g. during the night for home and during school hours for the school).

\subsection*{Mode Identification}
\label{sec:modeid}

As previously mentioned, the SENSg sensor is equipped with a Wi-Fi module, which scans the information of surrounding Wi-Fi hotspots (MAC addresses, Service Set Identifier, etc.) every 13 seconds. Each 13-second interval of collected information is called a working cycle. Once the data is uploaded to the server, the Wi-Fi hotspots information is converted to estimated latitude and longitude. As explained in \citep{Zhou2017}, a first cleaning process is carried out to reduce the noise in the raw data. Then, each package of 100 working cycles is considered as a single sample, and the transportation mode identification algorithm starts by classifying the sample into:
\begin{itemize}
\item Vehicle samples: Samples that belong to vehicle trips, \textit{or}
\item Non-vehicle samples: Samples that can be either walking or stationary.
\end{itemize}

After smoothing the data by a heuristic approach, vehicle samples and non-vehicle samples are processed separately. We refer the interested reader to \citep{Zhou2017} for a detailed description of the treatment of non-vehicle samples. In the following, our intention is to provide a brief overview of how vehicle samples are processed. The properties of this particular algorithm are crucial to this work, since we endeavor to estimate accurate measures of regret by taking into account the best clean subset of pre-processed information available.

The key idea behind the algorithm is to conduct type classification on segments instead of individual data samples. There are two main parts in this algorithm: first, catch segments and second, classify them. In the first part (\emph{catching}), the vehicle segment detection works by using information about a set of parameters such as the velocity and acceleration of the sample. Note that at this stage some segments can belong to walking or stationary modes. In the second part (\emph{classifying}), all detected vehicle segments are classified into one of the transportation modes. By following 17 segment-wise features such as the distance between segment's start/end location to the nearest bus/metro stop, etc. this goal can be reached. Hence, the vehicle trips can be properly classified since it is possible in each sample to know how close public transportation lines are, and how long/fast the vehicle segment is. Car mode trips are segments that do not fall into a metro or bus classification, as returned by a Random Forest classifier.

In short, the algorithm first catches the potential vehicle mode trips and then classifies each of them into riding a metro, riding a bus or riding a car. A high level accuracy of 85\% is guaranteed, indeed superior to similar recent studies (e.g. \citep{Sankaran2014, Shin2015} or \citep{Zhu2016}).

\subsection{New algorithms for empirical inefficiency and equilibration}

\subsection*{Estimating Socially Optimal Routing}

We now offer more details on the optimal trip duration. The Google Directions API was queried for the best route and minimal trip duration, for both private vehicle transportation as well as public transit.

For car users, the API was called in several different time periods to obtain different traffic conditions. Two of the collections were made on a Tuesday at both 7 and 8 am, reflecting a higher load on the Singapore road network. The last collection was performed on a Sunday at 7 am, which offers typically favorable traffic conditions.

For public transportation users, the API was called for a trip starting on a Tuesday morning at 7 am. The API does not take the traffic into account for transit requests, an issue minimized by the predictibility of the train transportation and the lower duration variability for bus services (as buses tend to come more frequently during peak hours).

To minimize the number of requests to the API, we have used the grid clustering method described in a previous subsection and queried for the best route between the cluster centroid and a school, if some of its students' homes were located inside the cluster. Some of the requests did not return satisfactory results, either due to a Not Found error or when repositioning the starting point of the trip too far from the student's home, and these were subsequently dropped in the analysis. We refer to Table \ref{tab:optimal} for the number of datapoints collected for the classes of users taking either public or private transportation.

\begin{table}
\centering
    \begin{tabular}{lccc}
        \toprule
        & \multirow{2}{*}{Number of points} & \multicolumn{2}{c}{Mean optimal trip duration} \\\cline{3-4}
        &                                   & Heavy (mins) & Light (mins) \\
        \toprule
        Private transportation & 6,833 & 19'30 & 14'06 \\
        \toprule
        Public transit & 8,345 & \multicolumn{2}{c}{35'39} \\
        \toprule
    \end{tabular}
    \caption{A few statistics on the collected optimal trips}
    \label{tab:optimal}
\end{table}

\subsection*{Clustering and Empirical Regret}

We focus on the duration of the students' morning trip from home to school. To quantify how well the Singapore routing system performs, we can obtain a lower bound of the total cost incurred by the students from the comparison between similar trips. More precisely, we divide the subjects in clusters and find for each cluster the student that reached school in minimal time.

Our clusters are indexed by 4 different variables:
\begin{itemize}
    \item \textbf{Geographical location \( l \):} Students living in the same neighborhood will be grouped together. We show in the following results obtained for different cluster of size $r$.
    \item \textbf{Time of departure \( t \):} It is not accurate to compare a student departing from home at 6 am with one starting at 8 am. We therefore group together students leaving on the same day within the same time frame, using a window size of 20 minutes.
    \item \textbf{Destination \( s \):} Students going to the same school are grouped together. In the case that two or more schools share the same location (e.g. a Primary and a Secondary school), students attending either one of them are added to the same cluster.
    \item \textbf{Mode of transportation \( m \):} We use two different sets of modes of transportation. In one version, we separate students using a car to go to school from those who take public transportation. In the second version, we also differentiate students whose primary mode is the bus versus those who use the train.
\end{itemize}

\begin{figure}
  \centering
  \includegraphics[width=0.6\linewidth]{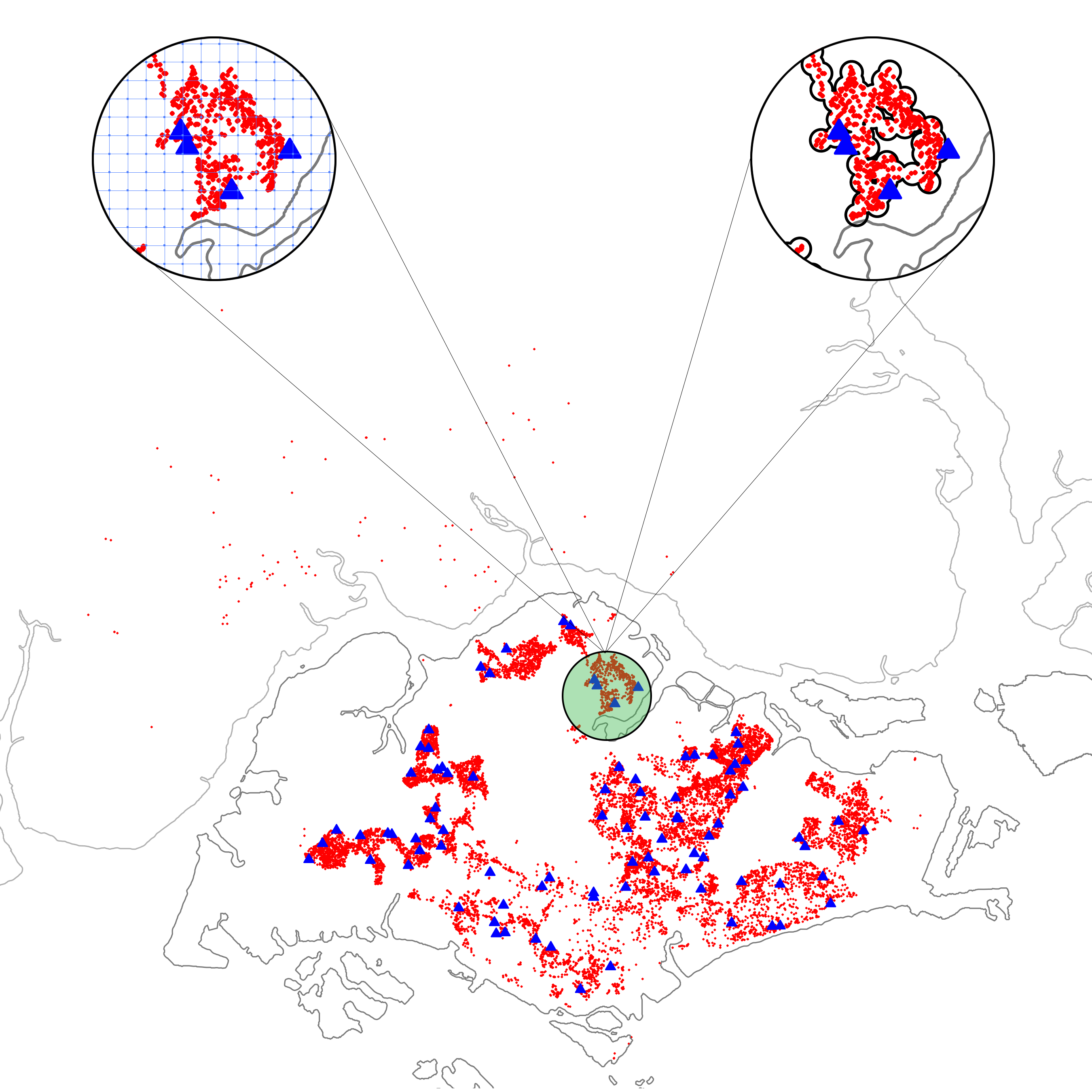}
  \caption{Home locations (red dots), school locations (blue triangles)  and spatial clustering methods.}
  \label{fig:clusters}
\end{figure}

To obtain the geographical locations $l$, two spatial clustering methods are implemented. In the first version, we find the smallest bounding box that contains all the home locations of the students. We divide this bounding box in cells of equal edge size \( r \), e.g. \( r = 400 \) meters, and assign to the same geographical clusters students with home locations inside of the same cell. This is a grid-based method that partitions the space into a finite number of cells from a grid structure. Its main advantage is its fast processing time. Sensitivity analysis measures are reported for different cell sizes.

The second version of the spatial clustering approach is based on a distance rule where all home locations of the students in the same cluster should be within $r$ meters of each other. This is a hierarchical clustering method using decision trees based in the geodesic distance matrix of all trips. This technique, although computationally more expensive, ensures that the distance rule holds for all the trips.

Figure \ref{fig:clusters} shows the visual comparison of the two different spatial clustering methods for $r=400$ meters for all 15,875 students considered in our study. The red dots represent the home locations and each dot corresponds to exactly one student. The blue triangles show the school locations of the 89 schools. It is interesting to note that some students have home location in Malaysia, who commute from Malaysia to Singapore daily for study. The grid-based method (top left) is a simple but efficient strategy, simply counting the points that fall in specific cells of the mesh. On the other hand, the distance rule approach (top right) can be visualized by circles of diameter equal to 400 meters. Inside each circle, the maximum distance between any two home locations is 400 meters. The algorithm optimizes a criterion function and the centroid of each spatial cluster can be easily identified, making it a powerful method to build the clusters. Recall that Figure \ref{fig:clusters} is presented only for visualization purposes since inside each spatial cluster (cells/circles), students might be mixed among different transportation modes and different destination schools.

\begin{figure}
	\centering
	\begin{minipage}[c]{0.46\linewidth}
        \centering
        \includegraphics[width=0.95\textwidth]{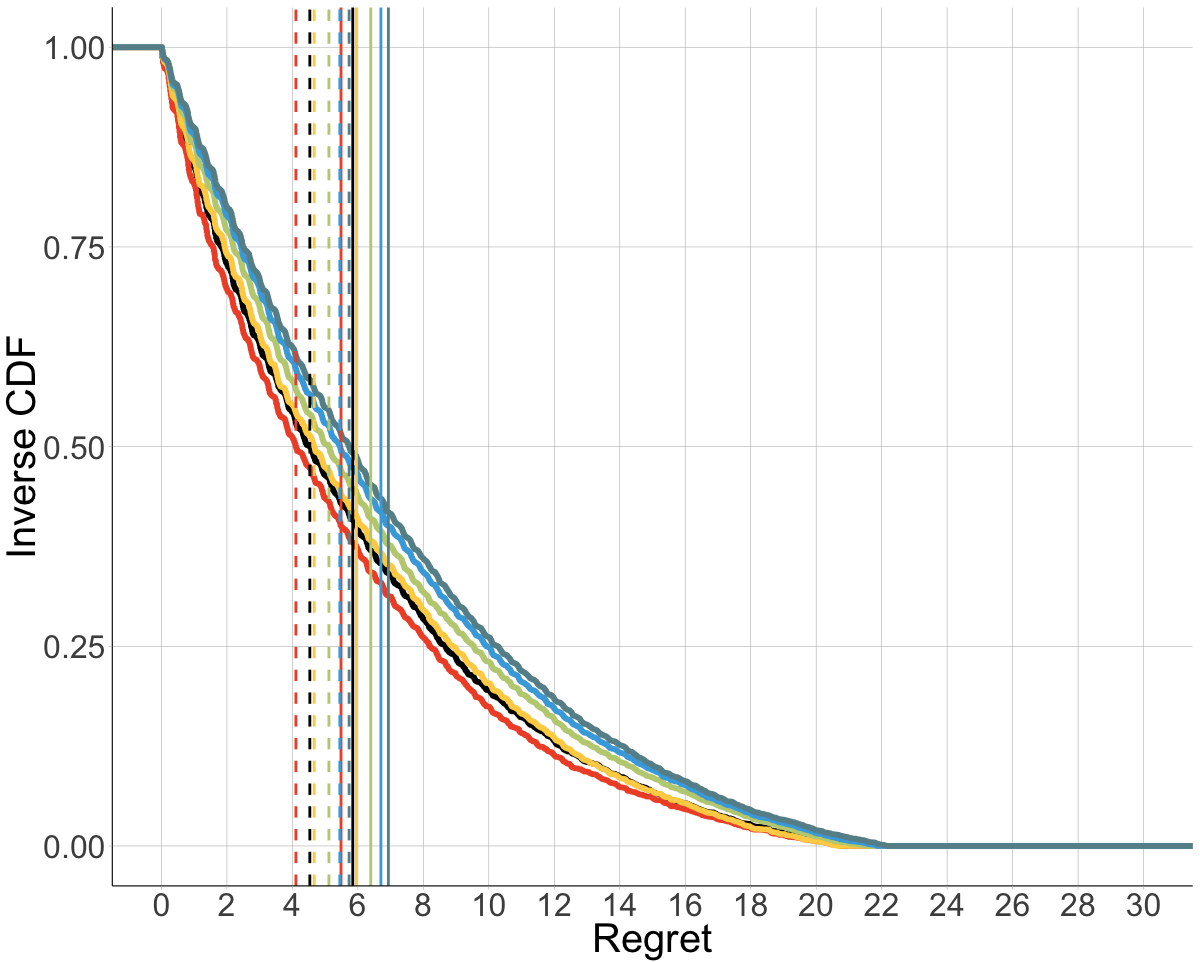}
	\end{minipage}
	\hspace{0.05\linewidth}
	\begin{minipage}[c]{0.46\linewidth}
		\centering
		\includegraphics[width=0.95\linewidth]{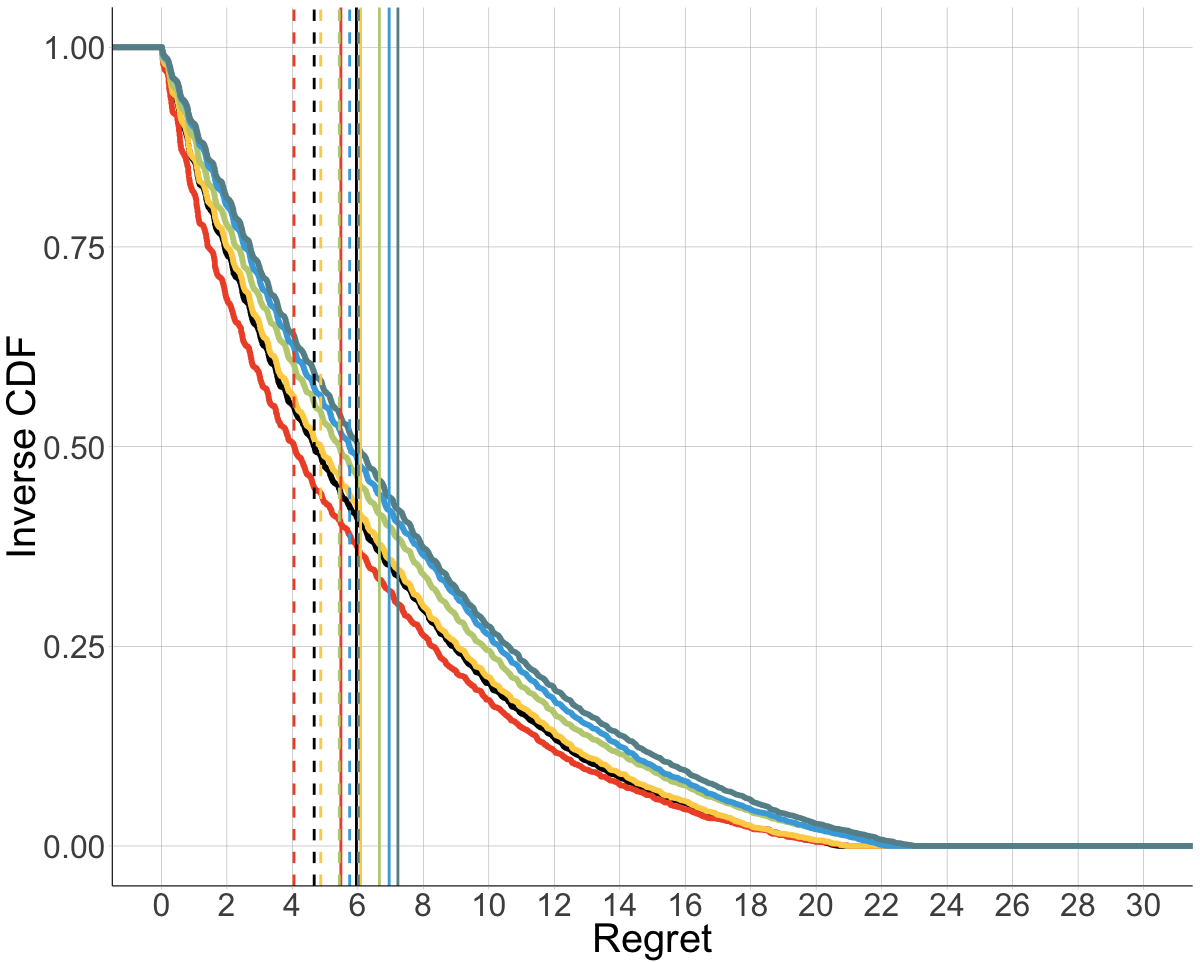}
	\end{minipage}
    \caption{The plots are similar to the one presented in the Findings section of the paper, with the difference that we are plotting the curves obtained from different clustering methods. \textit{Left:} We use as baseline the curve in black representing the results for the grid clustering with cell size 400 meters and plot in colors the curves obtained with different ball sizes (200, 400, 600, 800, 1000 meters, respectively red, yellow, green, blue and dark blue). Note that the means and medians increase as the size of the cluster grows. \textit{Right:} The baseline (in black) is now the curve for the ball clustering with 400 meters diameter and the colored curves correspond to the grid clustering with sizes (200, 400, 600, 800, 1000 meters, same colors).}
    \label{fig:regretcdfsens}
\end{figure}

We obtain a set of clusters \( \{ C_{l, t, s, m} \}_{l, t, s, m} \) where each \( C_{l, t, s, m} \) contains the trip durations \( t_{l, t, s, m}^i \) of students in the cluster. If several students belong to the same cluster, we find the student whose trip has the minimum duration among all trips in the cluster. We call this trip the \textit{baseline} \( t^b_{l, t, s, m} \), against which the remaining trips will be compared.

We next define the regret for student \( i \) in cluster \( C_{l, t, s, m} \) by
\[
    R^i_{l, t, s, m} = t^i_{l, t, s, m} - t^b_{l, t, s, m}
\]
Obviously the regret for the baseline student is zero, and non negative for everyone else. We are interested in seeing how large the deviations from the baseline can be, as a necessary condition for the system to be at equilibrium is that these deviations must be close to zero.

We plot in Figure \ref{fig:regretcdfsens} the empirical inverse cumulative distribution function of regret in minutes. To obtain the curve, we aggregate all measures \( R^i_{l, t, s, m} \) and remove the baseline trips of zero regret from the data. The inverse cumulative distribution shows for a point \( (x, \overline{F}(x)) \) the fraction of trips \( \overline{F}(x) \) that have regret greater or equal to \( x \), in minutes.

\subsection*{Consistency between trips}

Each student carries the sensor for up to 4 days in a week, allowing us to compare the morning trips taken by the same student between different days. Since the presence of noise in the sensor data and trip detection algorithm does not guarantee us that the whole week of experiment will be available, we filter out students for which only one morning trip is available. In our clean dataset, we have 15,875 individual students, out of which 9,352 have two or more trips logged in. On average, we have 2.44 trips per individual student. For these subjects, two analysis are carried out.

We first compare the modes of transportation selected by the student in the morning trips. Table \ref{tab:modeconsis} differentiates students that chose the same mode of transportation across all mornings from those that used multiple ones.

\begin{table}[!h]
\centering
    \begin{tabular}{lcccc}
        \toprule
        & Metro & Bus & Car & \textbf{Total} \\
        \toprule
        Number of students & 1,278 & 1,448 & 3,070 & 5,796 \\
        \toprule
        Number of students (bus and metro grouped) & \multicolumn{2}{c}{3,060} & 3,070 & 6,130 \\
        \toprule
    \end{tabular}
    \caption{Number of students using consistently the same mode of transportation}
    \label{tab:modeconsis}
\end{table}

In a second and more granular analysis, we compare the routing decisions of the student at road level. Our aim is to determine whether the student selects the same route consistently to reach school in the morning. To achieve this task, we need a distance \( d_{a, b} \) measuring the similarity between two sequences \( a= (a_i)_{i=1}^{n} \) and \( b = (b_j)_{j=1}^{m} \) of coordinates. If the sequences are close, we can conclude that the same route has been selected on two occasions.

The distance \( d_{a, b} \) is obtained with an estimation of the area enclosed by the polygon formed by the concatenation of \( a \) and \( b \). More formally, since  \( a_1 = b_1 = \text{ Home} \) and \( a_n = b_m = \text{ School} \), we consider the area enclosed by the polygon \( (a_1, a_2, \dots, a_n, b_{m-1}, b_{m-2}, \dots, b_1) \). The distance gets more precise as the number of datapoints logged along the trip increases, although it is still possible to achieve good results for very sparse trips.

\begin{figure}[h!]
	\centering
	\begin{minipage}[c]{0.46\linewidth}
		\centering
		\includegraphics[width=0.95\linewidth]{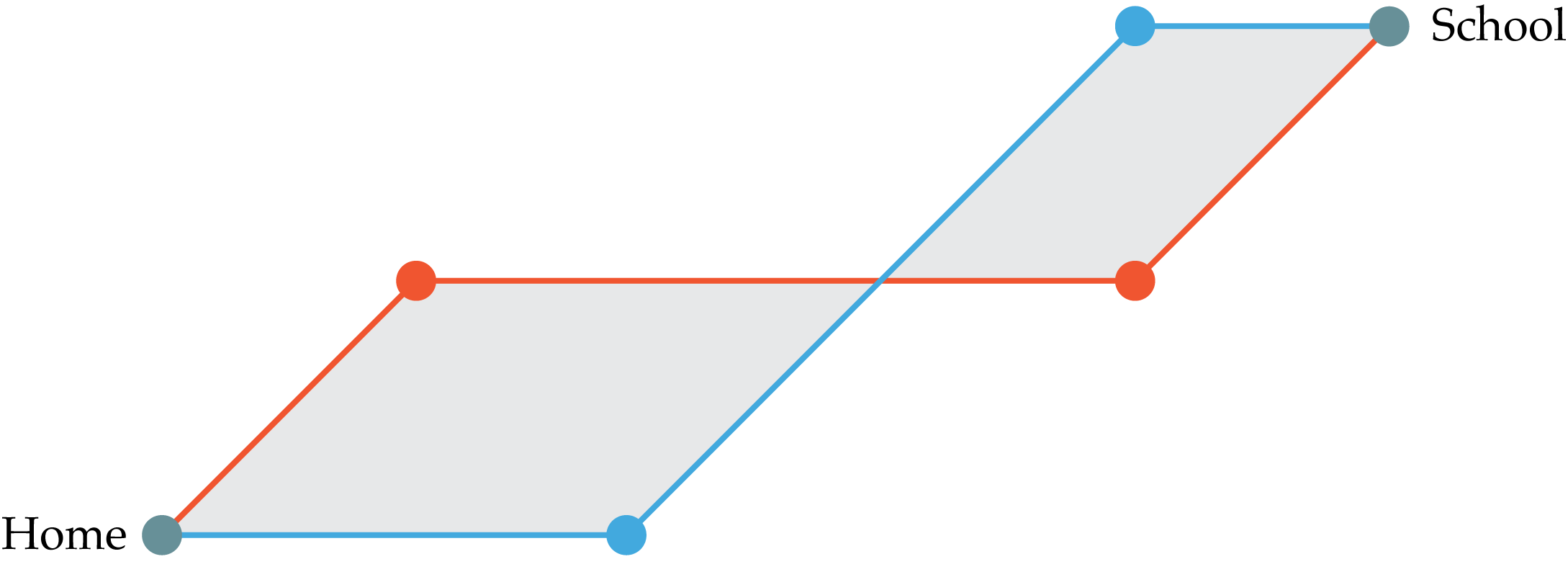}
		\caption{Two trips (red and blue) are plotted, with endpoints in dark blue. The distance between the trips is measured by the area of the darkened surface.}
		\label{fig:polygontrip}
	\end{minipage}
	\hspace{0.05\linewidth}
	\begin{minipage}[c]{0.46\linewidth}
		\centering
		\includegraphics[width=0.95\linewidth]{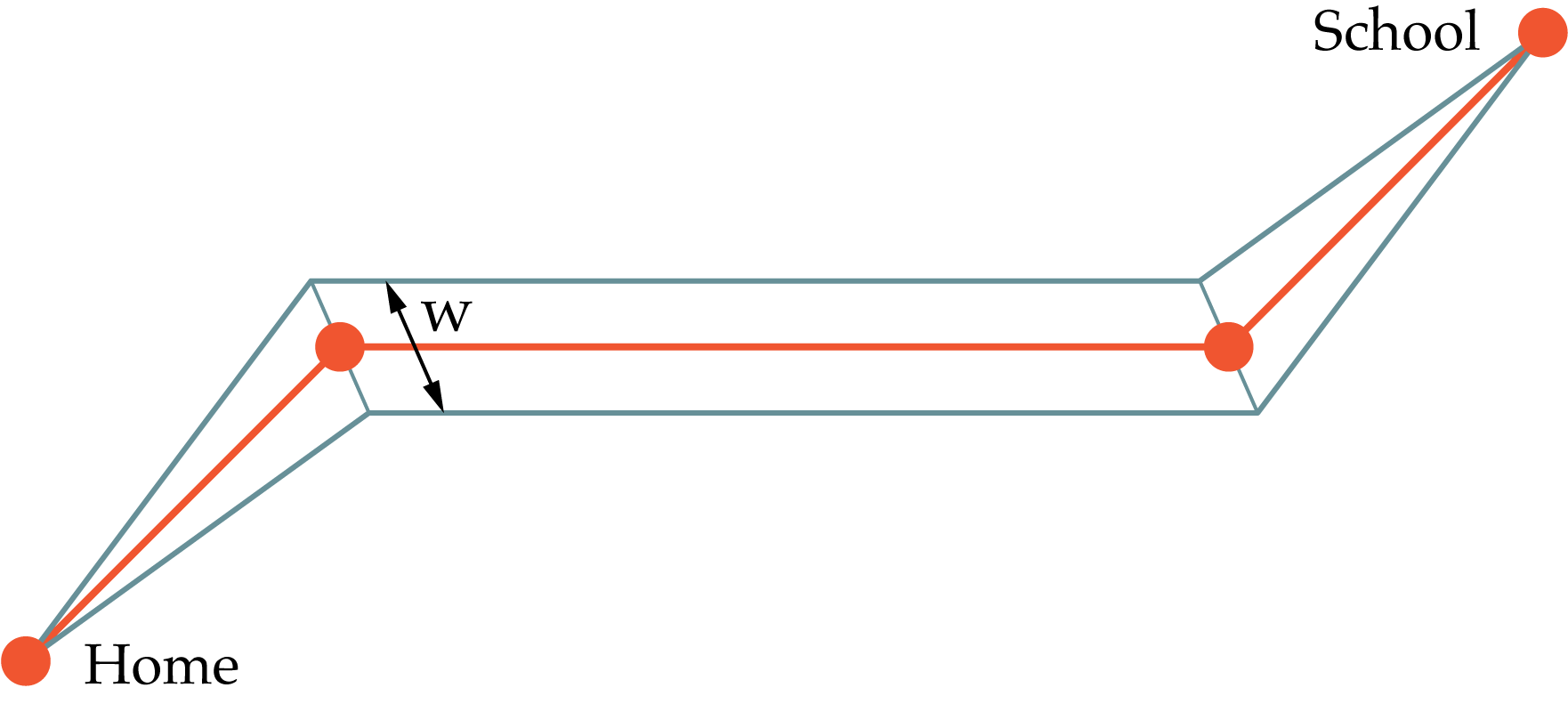}
		\caption{A four-point trip is plotted in red. The outer contour is obtained by fixing a band width \( w \).}
		\label{fig:outercontour}
	\end{minipage}
\end{figure}

We need a criterion to decide whether the previously obtained area is sufficiently small for the two sequences of coordinates to be considered consistent. To this end, we construct the outer contour of each sequence \( a \) and \( b \), defined by a polygon containing the sequence of coordinates. Intuitively, we construct a band around the stream of locations, and the area of that band allows us to determine what constitutes an acceptable deviation. We show in Figure \ref{fig:outercontour} a representation of the outer contour for a trip with 4 points. Given \( d^o_a \) and \( d^o_b \), respective areas of the outer contour for \( a \) and \( b \), we use the following criterion to classify the routes \( a \) and \( b \) as consistent:

\[
    d_{a, b} < \frac{d^o_a + d^o_b}{2} \Rightarrow a \text{ and } b \text{ are consistent.}
\]
The average is taken to ensure that both trips are considered equally in our criterion. Indeed, without the average, the algorithm may not be able to recognize a small deviation if only the outer contour of the shortest path appears on the right-hand side.

A parameter \( w \) controls the width of the band. If we set \( w \) to a value that is too large, we run the risk of incorrectly classifying different trips as consistent. On the other hand, a \( w \) that is too small may mark as non-consistent trips that make use of the same route. We have set the value of \( w \) by creating negative examples, comparing a trip with translated versions of the same sequence of coordinates. Visual analysis further confirmed the validity of its choice.

\section{Connections to Other Work}
{\bf Price of Anarchy for Real World Networks.} One earlier paper tangentially connected to computing the empirical PoA of congestion games is
\citep{buriol2011smoothed}. This is a theoretical paper that provides PoA bounds for  perturbed versions of congestion games.
As a test of their techniques,  they heuristically approximate the PoA  on  a few benchmark instances of traffic networks available for academic research from the Transportation Network Test Problems~\citep{BarGera} by running the Frank-Wolfe algorithm on them. No experiments were performed and no measurements were made. Naturally, this approach cannot be used to test PoA predictions, since it presumes that PoA reflects the worst case possible performance and then merely tests where do these constants lie for non-worst case routing networks.

In effectively parallel independent work \citep{zhang2016price, zhang2016data} focused on quantifying the inefficiencies incurred due to selfish behavior for a sub-transportation network in Eastern Massachusetts, US.
They use a dataset containing time average speed on road segments and link capacity in their transportation sub-network.
The authors estimate daily user cost functions as well as origin-destination demand by means of inverse optimization techniques using this dataset.
From this formulation they compute estimates of the PoA, whose average value is shown to be around $1.5$.
In contrast to their approach our dataset contains detailed individual user information, which allows for estimates not only of systemic performance but also of individual optimality (e.g. regret) as well as test to what extent is the system indeed near stasis (i.e. in equilibrium).


{\bf Algorithmic Game Theory and Econometrics.} Recently there has been a surge of interest in combining techniques from algorithmic game theory with the traditional goals of econometrics \citep{syrgkanis2015algorithmic}.
In \citep{Nekipelov:2015:ELA:2764468.2764522} the authors developed theoretical tools for  inferring  agent valuations from observed data in the generalized second price auction without relying on the Nash equilibrium assumption, using behavioral models from online learning theory such as regret-minimization.  They applied their techniques on auction data to test their effectiveness.
In \citep{hoy2015robust} the authors provided tools for estimating the empirical PoA of an auction. The empirical PoA is defined as the worst case efficiency loss of any auction that could have produced the data, relative to the optimal. 
 However, the two domains, auctions and routing games, are quite distinct and each poses a totally distinct set of challenges. Also, in our setting the problem of translating data streams to game theoretic concepts  adds a rather nontrivial layer of complexity. For example, even identifying the action chosen by each agent, i.e. their route is tricky as it requires to robustly map a noisy stream of transportation data into a discrete object, a path in a graph.

{\bf Transportation Science and Game Theory.} The Braess paradox famously asserts that adding a road to a congested traffic network could increase the overall journey time, and in effect, increase the PoA of the network. It is also known to be indeed a prevalent phenomenon in real traffic networks~\citep{steinberg1983prevalence}. In \citep{bagloee2014heuristic}, the authors examine heuristics to identify roads in traffic networks whose removal would  improve the overall system performance, applying the Braess paradox in the reverse direction. This is an interesting example where a PoA-related phenomenon inspires a possible improvement to a transportation problem.

{\bf Singapore's National Science Experiment.}
NSE is a science experiment carried out by Singapore students.
The experiment involves students carrying a specially designed sensor to collect data on their daily travel as well as data from the environment. This data is transferred wirelessly to a central online portal from which the students can log in to view the results, including the aggregated data of students from all over Singapore. The design of the sensor is explained here \citep{Wilhelm2016}. This work constitutes the top (application) layer of a deep layered architecture including
 state-of-the-art inference methods~\citep{Monnot2016, Wilhelm2017, Zhou2017} and going all the way down to optimized network protocols and in-house designed hardware that have been  continuously improved upon over the span of the last three years.

\section{Conclusion and Open Questions}
This is hopefully not the end but the beginning of a thorough experimental investigation into the rich literature of price of anarchy in routing games. Do these estimates remain robust to increases of the size of the sample? What about different cities around the globe? Is Singapore's routing efficiency an artifact of its traffic topology and local cultural norms or are these estimates relatively robust across many cities and cultures?  Putting theoretical predictions under detailed experimental scrutiny holds the promise of surprising insights that may fuel more informed theoretical investigations. With this work, we hope to kickstart this positive feedback cycle.

\section*{Acknowledgements}
The authors would like to thank the National Science Experiment team at SUTD for their help: Garvit Bansal, Sarah Nadiawati, Hugh Tay Keng Liang, Nils Ole Tippenhauer, Bige Tunçer, Darshan Virupashka, Erik Wilhelm and Yuren Zhou. The National Science Experiment is supported by the Singapore National Research Foundation (NRF), Grant RGNRF1402.

Barnab\'e Monnot would like to acknowledge a SUTD Presidential Graduate Fellowship.
Georgios Piliouras would like to acknowledge
	SUTD grant SRG ESD 2015 097 and MOE AcRF Tier 2 Grant  2016-T2-1-170.
		Part of the work was completed  while Barnab\'e Monnot and Georgios Piliouras  were visiting scientists at the Simons Institute for the Theory of Computing (Economics and Computation Semester, Fall '15).


\bibliographystyle{apalike}
\bibliography{references,sigproc2,sample-bibliography}

\end{document}